\documentclass[aps,notitlepage,twocolumn,prl,tightenlines,footinbib,superscriptaddress,showpacs,floatfix]{revtex4-1} 

\usepackage{amsmath,amssymb,amsfonts,latexsym}
\usepackage[mathcal]{euscript}
\usepackage{graphicx}
\usepackage{epsfig}
\usepackage{color}
\usepackage[activate=normal]{pdfcprot}  

\usepackage{cases}

\usepackage{psfrag}
\usepackage{pifont}
\usepackage[caption = false]{subfig}

\newcommand{\eccm}{Gulliver Lab UMR CNRS 7083, ESPCI Paris, PSL Research University, 75005 Paris, France}
\newcommand{\phenix}{Sorbonne Universit\'e, CNRS, Laboratoire PHENIX, UMR CNRS 8234, 75005 Paris, France}

\newcommand{\be}{\begin{equation}}
\newcommand{\ee}{\end{equation}}
\newcommand{\ben}{\begin{equation*}}
\newcommand{\een}{\end{equation*}}
\newcommand{\ba}{\begin{eqnarray}}
\newcommand{\ea}{\end{eqnarray}}

\newcommand{\sgn}[0]{\mathrm{sgn}}
\newcommand{\ex}[1]{\mathrm{e}^{#1}}

\newcommand{\od}[1]{\begingroup\color[rgb]{0,0,0}#1\endgroup}
\newcommand{\odbis}[1]{\begingroup\color[rgb]{0,0,0}#1\endgroup}

\begin{document}
\graphicspath{{./Figs/}}

\title{\odbis{Speed-Dispersion Induced Alignment : \\ a 1D model inspired by swimming droplets experiments.}}

\author{Pierre Illien}
\affiliation{\eccm}
\affiliation{\phenix}
\author{Charlotte de Blois}
\affiliation{\eccm}
\author{Yang Liu}
\affiliation{LMIS2, Ecole Polytechnique Fédérale de Lausanne, CH-1015 Lausanne, Switzerland}
\author{Marjolein N. van der Linden}
\affiliation{\eccm}
\author{Olivier Dauchot}
\affiliation{\eccm}

\date{\today}

\begin{abstract}
We investigate the collective dynamics of self-propelled droplets, confined in a one dimensional micro-fluidic channel. On one hand, neighboring droplets align and form large trains of droplets moving in the same direction. On the other hand, the droplets condensates, leaving large regions with very low density.  A careful examination of the interactions between two "colliding" droplets demonstrates that \odbis{local alignment} takes place as a result of the interplay between \od{the dispersion of their speeds} and the absence of Galilean invariance. \odbis{Inspired by} these observations, we propose a minimalistic 1D model of active particles \odbis{reproducing such dynamical rules} and, combining analytical arguments and numerical evidences, we show that the model  exhibits a transition to collective motion in 1D for a large range of values of the control parameters. Condensation takes place as a transient phenomena which tremendously slows down the dynamics, before the system eventually settles  into a homogeneous aligned phase.
\end{abstract}

\maketitle

Collective dynamics in systems of active particles have been the topic of a fantastic amount of work~\cite{Vicsek:2012ty,Marchetti:2013bp,Bechinger:2016cf}. Both the transition to collective motion~\cite{Chate:2008isb,Vicsek:2012ty,Solon:2013vr,Peshkov:2014un} and the Motility Induced Phase Separation (MIPS)~\cite{Cates:2015ft}  are now well understood. The interplay of alignment and crowding effects have been investigated more recently~\cite{Farrell:2012ks,SeseSansa:2018wl,Shi:2018tp,vanderLinden:2019kd,Lozano:2019er}. The majority of these studies have however been conducted in two dimensional space and much less is known about active systems in one dimension.

Yet the physics of active system in 1D is relevant as soon as the confinement breaks the continuous symmetry of the order parameter describing collective motion. It is the case in systems of highly confined bacteria ~\cite{Biondi:up}, pedestrians~\cite{Jelic2012} or molecular motors~\cite{Chou2011}. Also, from a more theoretical point of view, 1D systems often exhibit peculiar dynamics~\cite{lieb2013mathematical,privman2005nonequilibrium}, resulting from the presence of strong correlations, as exemplified by single-file diffusion~\cite{Arratia1983,Harris1965}, \od{1D inelastic dynamics~\cite{Du:1995kw,BenNaim:1999gj,Baldassarri:2018id}} or 1D exclusion processes~\cite{Mallick2015}. 

In the context of active matter, kinetic theory results~\cite{Peshkov:2014un,Manacorda:2017eia} or generic arguments relying on the non conservation of momentum~\cite{Lam:2015jr} do not easily generalize in 1D because of the discrete symmetry of the polar order parameter. Despite this limitation, a few models were put forward to describe 1D active systems. The \lq\lq active Ising model\rq\rq~\cite{Solon:2013vr,Solon:2015he}, a stochastic lattice gas model, has been decisive in our current understanding of the transition to collective motion in terms of a bona fide liquid-gas phase transition. It however does not include steric interactions. Conversely, several models were proposed to describe the clustering transition in assemblies of excluding run-and-tumble particles, but do not include alignment~\cite{Soto2014,Slowman2016,Slowman2017,Locatelli:2015it}.
Finally, hydrodynamic limits were derived exactly provided that the different processes have appropriate scalings~\cite{Kourbane-Houssene2018a}, but this approach still lacks the combined effect of volume-exclusion and alignment.

\begin{figure}[t!]
\vspace{-0mm}
\includegraphics[width=0.98\columnwidth,trim = 0mm 0mm 0mm 0mm, clip]{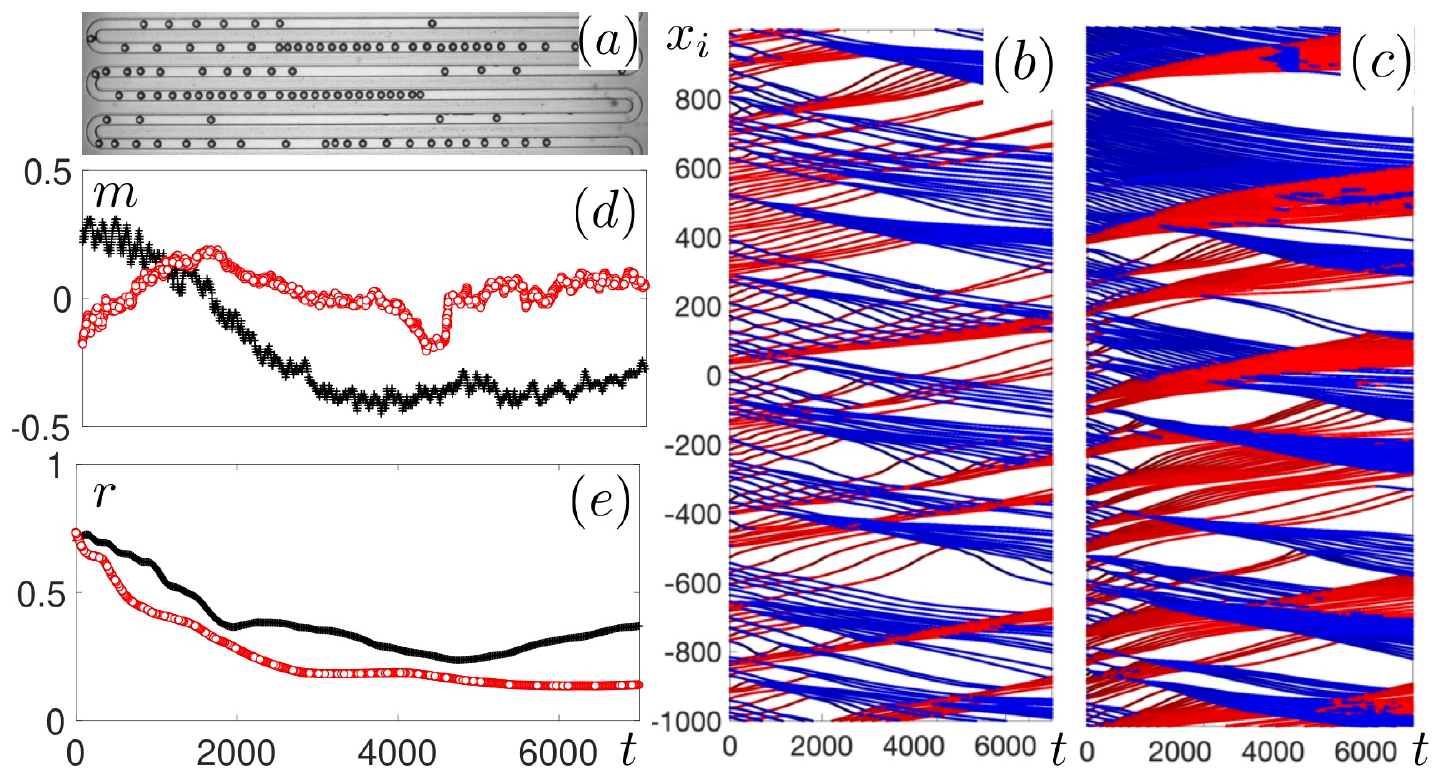}
\vspace{-0mm}
\caption{{\bf Collective dynamics of swimming droplets in a micro-fluidic serpentine} (a) Part of the set-up; (b,c) Spatio-temporal diagram of the dynamics for packing fraction $\phi = Na/L = [0.082; 0.122 ]$ (blue and red colors code for the direction of motion; a trajectory changes color when a droplet reverses its direction; time in $s$ and space in droplet diameter). (d,e) Polarisation $m$ and participation ratio $r$ (see main text) vs. time for $\phi = 0.082$ (black) and $\phi = 0.122$ (red).}
\label{fig:dst}
\vspace{-5mm}
\end{figure}
In this Letter we first report on the observation of collective alignment and spatial condensation in a one dimensional system of swimming droplets (Fig.~\ref{fig:dst}):  'trains of droplets' spontaneously form and move coherently. To the best of our knowledge, this is the first experimental realization and observation of the onset to collective motion in a one-dimensional active system. The analysis of the short time dynamics, resulting from the interaction of two droplets, reveals that the \odbis{local} alignment results from the interplay between \od{the dispersion of the droplet speeds} and the absence of Galilean invariance.  \odbis{Wether such a mechanism leads to a large scale transition to collective motion, especially in 1D, is far from obvious}. We propose a minimal model based on these observations, and, combining analytical arguments and numerical evidences, we show that \odbis{a system of active particles reproducing the observed local dynamical rules} exhibits a transition to collective motion for a large range of values of the control parameters. Condensation takes place as a transient phenomena which tremendously slows down the dynamics, before the system eventually reaches a homogeneous aligned phase. \odbis{Let us stress that our model does not aim at being a model of swimming droplets. Rather the swimming droplets experiments have pointed at an intriguing phenomena, raising a conceptual interrogation, which we answer to}.

The experimental system is composed of $N\in[50 - 500]$ swimming water droplets~\cite{Izri:2014fv} of diameter $a=200$ $\mu$m, confined in a micro-fluidic square channel of section $h^2 = 200 \times 200$ $\mu \text{m}^2$ and length $L$ [Fig.~\ref{fig:dst}(a)], filled with a surfactant-in-oil solution, with concentration far above the critical micellar concentration.
The swimming motion of the water droplets results from the combination of two ingredients~\cite{Michelin:2013gv,Izri:2014fv}. First, the system is away from its physico-chemical thermodynamic equilibrium: a slow but steady flux of water takes place from the droplet to the surfactant micelles. Second, the resulting isotropic concentration field of swollen micelles is unstable against an infinitesimal flow disturbance: concentration gradients parallel to the droplet surface spontaneously form, and induce Marangoni stresses and phoretic flows, which are in turn responsible for the motion of the droplet.
The droplets are introduced in a micro-fluidic serpentine using standard micro-fluidic techniques. Once all external fluxes are interrupted, their swimming motion is tracked with a high resolution CCD camera to obtain their curvilinear abscissa $x_i(t)$, $i\in[1,N]$, corrected from residual drifts, displayed in Fig.~\ref{fig:dst}(b,c). (see also Supp. Mat.~\cite{suppmat}).

\od{Initially each droplet picks up a random direction. It swims straight, \emph{with no tumbling}, until it interacts with another droplet.} Once the droplets interact, trains of droplets moving in the same directions form, pointing at the presence of an alignment mechanism. These trains of droplets interact with each other leading to the rich spatio-temporal dynamics reported in Fig.~\ref{fig:dst}(b,c). Note that the droplets in a train are close to but do not touch each other (see Movie-1 of Supp. Mat.~\ref{suppmat}). 
We define the average polarisation $m(t)=\left< s_i(t) \right>$, \od{which characterizes the amount of orientational ordering} the and the participation ratio $r(t)=\frac{\left<d_i\right>^2}{\left<d_i^2\right>}$, \od{which quantifies the homogenitiy of the spacing between droplets}, where $s_i(t)=\pm 1$ codes for the orientation of the droplet motion,  $d_i(t) = x_{i+1}(t) - x_i(t)$ and $\left<\bullet\right> = \frac{1}{N}\sum_i (\bullet)$. While the condensation is nicely confirmed by the monotonous decrease of $r$ [Fig.~\ref{fig:dst}(e)], the dynamics of the polarisation is far more complex [Fig.~\ref{fig:dst}(d)]: large and slow variations of $m$ suggest large scale reorganizations of the aligned domains, but no clear tendency towards a global polarization on the time scale of the experiment.

\begin{figure}[t!]
\vspace{-0mm}
\includegraphics[width=0.95\columnwidth,trim = 0mm 0mm 0mm 0mm, clip]{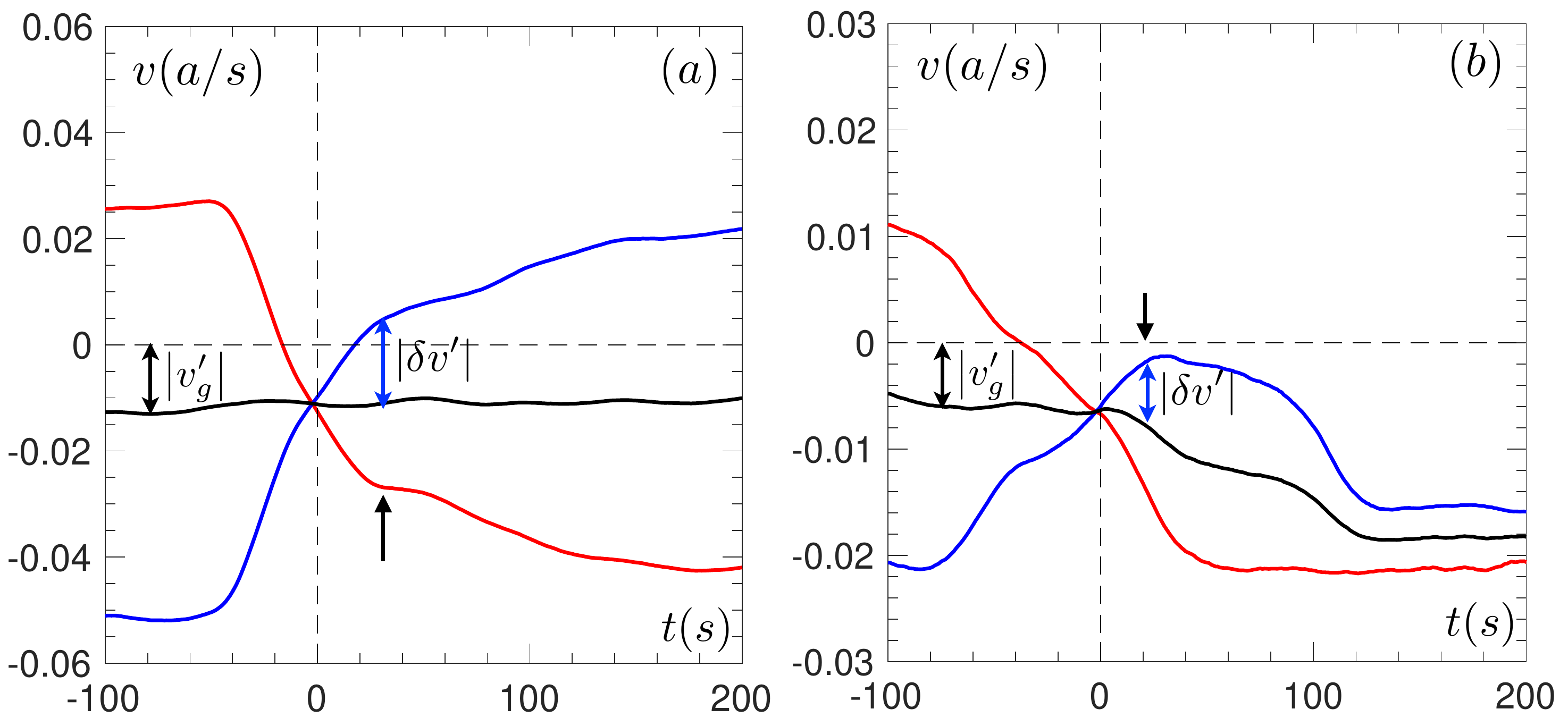}
\vspace{-3mm}
\caption{{\bf Binary interactions :} velocities $v_1, v_2$ of two interacting droplets (one red, one blue line); and center of mass velocity $v_g$ (black line) for two different collisions; the vertical dashed line indicates the time of collision; the arrow indicates the time at which relaxation towards the nominal velocity takes place: \od{(a) $\left|\delta v'\right|\!>\!\left| v'_g\right|$:  the droplets bounce against each other; (b) $\left|\delta v'\right|\!<\!\left| v'_g\right|$ : the droplets motion align.}}
\label{fig:coll}
\vspace{-5mm}
\end{figure}

Let us now focus on the interaction between pairs of droplets [Fig.~\ref{fig:coll}].  When two droplets moving in opposite direction encounter, their speeds, just after the collision, are smaller than before, suggesting to describe the interaction as an effective inelastic collision. \odbis{On the basis of a statistical analysis of several hundreds of collisions,} we could evaluate a restitution coefficient $\alpha\simeq 0.4\pm 0.15$. The droplets being active, their speeds then relaxe towards their nominal active speed $v_0$. This relaxation process is slow and is generally interrupted by a new interaction event. Therefore, even for the dilute systems considered here, the \od{speed} of the droplets are strongly dispersed and essentially never relaxed to $v_0$. \od{The standard deviation of the speed is typically $\sigma_v\simeq 0.3\,v_0$.} \od{As a result, in most collisions,} incoming droplets have different speeds, and the velocity of the center of mass is generically non zero. [Fig.~\ref{fig:coll}(a)] and [Fig.~\ref{fig:coll}(b)] illustrate the two possible scenario taking place in such a context.  \od{Denoting with a prime the velocities after collision, one sees that, in [Fig.~\ref{fig:coll}(a)] (resp.  [Fig.~\ref{fig:coll}b]), $\left|\delta v'\right| = \left|v'_2 - v'_1\right|/2$ is larger (resp. smaller) than $\left| v'_g\right| = \left| v'_2+ v'_1\right|/2$. Therefore, the velocities of the two droplets, just after collision and before relaxation, have opposite (resp. equal) signs. At that point starts the active relaxation towards $v_0$. When the velocities after collision have opposite sign, everything takes place as if the two droplets had bounced on each other. On the contrary, when the velocities after collision have equal sign, the relaxation leads to a net alignment. Note the importance of studying the collisions in the reference frame of the lab :  it is the combination of speed dispersion and  the absence of Galilean invariance, which is at the root of the alignment mechanism.}

\begin{figure*}[t!]
\vspace{-0mm}
\includegraphics[width=0.98\textwidth,trim = 0mm 0mm 0mm 0mm, clip]{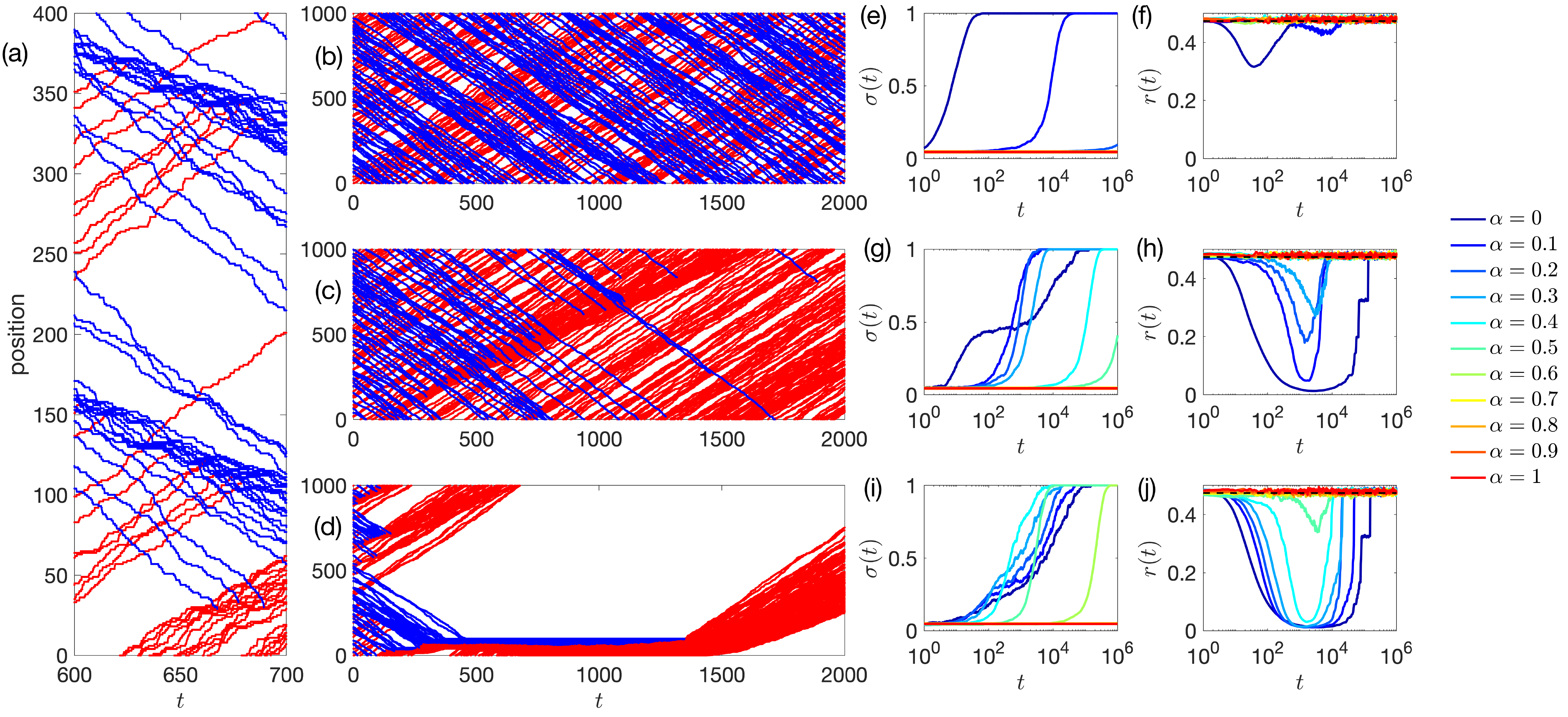}
\vspace{-2mm}
\caption{{\bf Spatiotemporal dynamics.} Numerical simulations of 100 particles on 1000 sites ($\phi=0.1$). (a) Transients on the same time and length scales as the experiment with a relaxation rate $\gamma=1$ and an inelastic coefficient $\alpha=0.43$ (zoom-in on the whole trajectory). Particles with positive (resp. negative) velocity are colored in red (resp. blue) : a trajectory changes color when a droplet reverses its direction. (b, c, d) Typical trajectories observed for $\alpha=0.2$ and $\gamma = 100, 10, 1$ (from top to bottom). (e-j) Magnetization and participation ratio (averaged over 100 realizations) as a function of time for different values of $\alpha$ and the relaxation rates corresponding to the central panels: $\gamma=100$ (e-f), $\gamma=10$ (g-h) and $\gamma=1$ (i-j).} 
\label{fig:dstnum}
\vspace{-3mm}
\end{figure*}

\odbis{Wether the above minimal ingredients are sufficient to observe a transition to  collective alignment is far from obvious. First, regarding the present system of droplets, the hydrodynamics and the non linear coupling to diffusion may well not reduce to such a simple description. 
Second, from a more general perspective, long range order  in one dimension is prone to be destroyed by fluctuations. It is therefore of interest to investigate the onset of collective motion within a minimal model, which share the same local dynamics as the one reported above.}
The stochastic model we propose is as follow. $N$ particles evolve on a 1D lattice of $L$ sites with periodic boundary conditions (site $L+1$ $\equiv$ site 1), the density is $\phi=N/L$.\ The $N$ particles are initially placed at random on the lattice with the condition that there can only be one particle per site. The initial velocity of each particle $v_i(0)$ is drawn uniformly from the interval $[-1,1]$ and  the centre-of-mass velocity is subtracted from the initial velocity of each particle.  At time $t$, the velocity obeys the exponential relaxation law:
\begin{equation}
v_i(t) = \sgn(v_{c,i})v_0 + [v_{c,i}-\sgn(v_{c,i})v_0 ]\ex{-\gamma(t-t_{c,i})},
\end{equation}
where $t_{c,i}$ is the last collisional time for particle $i$, $v_{c,i}$ its velocity right after that collision, and $\gamma$ is the relaxation rate.. 

At each time step, a particle $i$ at site $x_i$ is chosen at random. Say $v_i>0$; three situations can occur depending on the occupation of site $x_i+1$: (i) if site $x_i+1$ is free, particle moves to this location with probability $\left| v_i(t)\right|/v_0$; (ii) if site $x_i+1$ is occupied but $v_{i+1}\ge v_i>0$, nothing happens; (iii) if site $x_i+1$ is occupied and $v_i>v_{i+1}$, a collision occurs with probability $(v_i-v_{i+1})/v_0$, following the inelastic collision rule:
\begin{equation}
\label{collisionrule}
\begin{pmatrix}
      v'_i    \\
        v'_{i+1}
\end{pmatrix}
=
\begin{pmatrix}
  \frac{1-\alpha}{2}    & \frac{1+\alpha}{2}   \\
  \frac{1+\alpha}{2}    & \frac{1-\alpha}{2}   
\end{pmatrix}
\begin{pmatrix}
      v_i    \\
      v_{i+1}  
\end{pmatrix},
\end{equation}
where $\alpha \in [0,1]$ is the inelastic coefficient. The nominal velocity $v_0=1$ and the lattice spacing $a=1$ so that time is measured in units of $a/v_0$. On average, each particle only moves every $N$ time steps, so that one time unit corresponds to $N$ Monte-Carlo time steps.

The model \od{encodes} the binary interactions observed experimentally, namely that alignment takes place as soon as $\left|\delta v'\right| = \left|v'_i - v'_{i\pm 1}\right|/2$ is smaller than $\left| v'_g\right| = \left| v'_i + v'_{i\pm 1}\right|/2$.  The first key observation, it that the transient collective dynamics reproduces \od{qualitatively} the experimentally observed formation of trains of droplets and the way they interact [Fig~\ref{fig:dstnum}(a)]. The effective microscopic rules we designed are thus sufficient to mimic the experimental system and we can now take advantage of the numerical simulations to characterize the long time dynamics in a wide range of parameters. We observe three possible evolutions of the initial disordered state. For large $\gamma$ or $\alpha \simeq 1$, particles relax rapidly to the nominal velocity between collisions and behave effectively like elastic particles [Fig.~\ref{fig:dstnum}(b)]. In this  situation, the  absolute polarization averaged over initial conditions and noise realizations, $\sigma = \overline{\left|m\right|}$, remains close to 0 [Fig. \ref{fig:dstnum}(e-g-i)-red to orange curves] and the participation ratio is close to its homogeneous value $(1-\phi)/(2-\phi)$ [Fig.~\ref{fig:dstnum}(f-h-j)-red to orange curves]. For lower values of  $\gamma$, or $\alpha$, the speed lost, resulting from the inelastic collisions, is not immediately compensated; dispersion of speeds set in and alignment emerges. The transition to collective motion may take two different routes. For intermediate values of $\gamma$ and $\alpha$, the polarization increases rapidly from $0$ to $1$ while the system remains spatially homogeneous. An example of such an evolution is shown on Fig.~\ref{fig:dstnum}(c). For even smaller values of  $\gamma$ or $\alpha$, the transition to the aligned state is interrupted by the formation of large clusters made of two groups of particles (one with positive velocity, one with negative velocity), before it eventually resumes, when the boundary between the two groups reaches the edge of the cluster [Fig.~\ref{fig:dstnum}(d)]. This transient clustering translates into a significant drop of the participation ratio, as observed on [Fig. \ref{fig:dstnum}(h,j)]. 
\begin{figure}[t!]
\vspace{-0mm}
\includegraphics[width=0.98\columnwidth,trim = 0mm 0mm 0mm 0mm, clip]{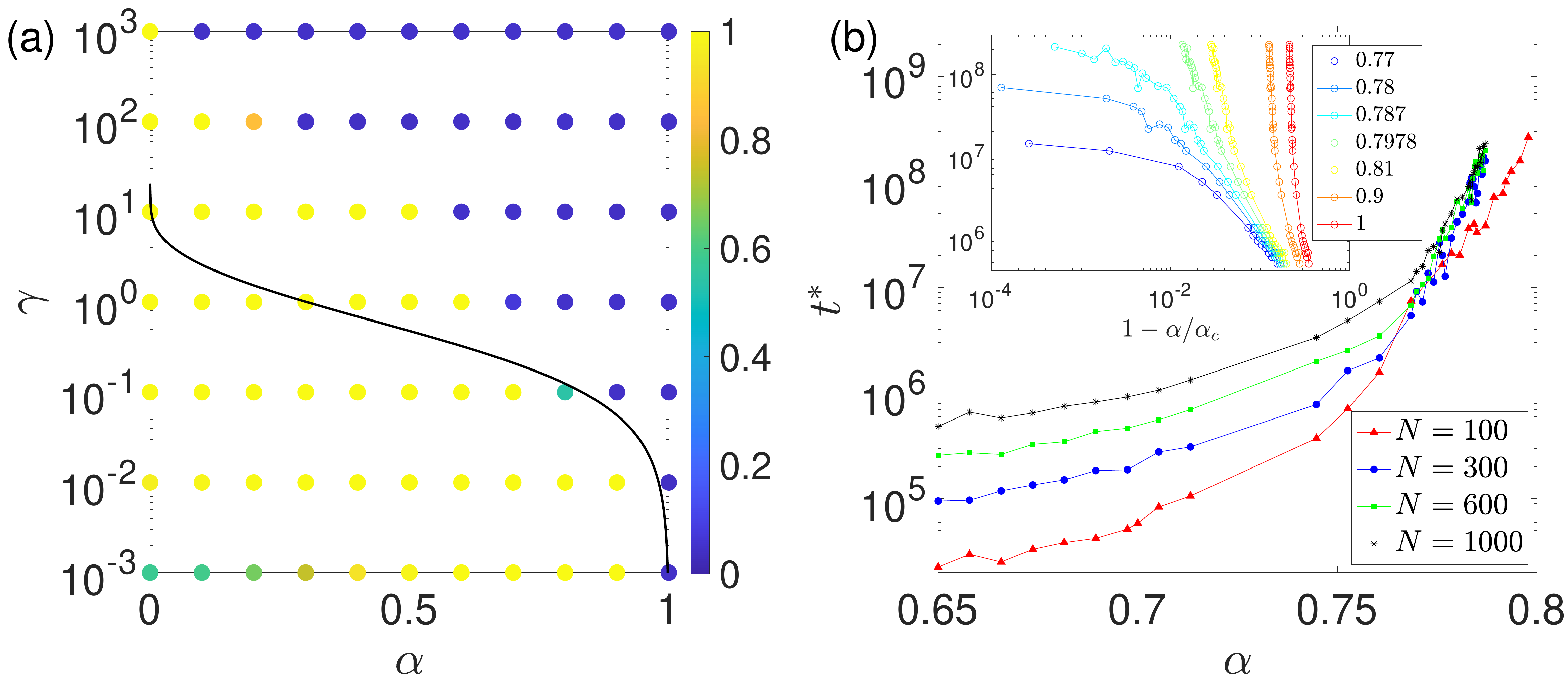}
\vspace{-0mm}
\caption{{\bf Phase diagram.} (a) Parameter space $(\alpha, \gamma)$ color coded by the magnetization computed at $t_\text{max}=10^7$, averaged over 100 realizations, and $\phi=0.1$ (100 particles on 1000 sites). The solid line is a plot of $\gamma_c(\alpha)$ with $d=1$ (Eq. \eqref{Gamma_infinity_main}). (b) Transition time $t^*$ (at which $\sigma=0.1$ for the first time) vs. $\alpha$ for $\gamma = 0.1$ and $\phi=0.1$ (inset : $t^*$ vs. $(\alpha_c-\alpha)/\alpha_c$ for $N=1000$ and different choices of $\alpha_c$ as given in the legend).}
\label{fig:phasediag}
\vspace{-3mm}
\end{figure}

Fig.~\ref{fig:phasediag}(a) displays the corresponding phase diagram, obtained from the values of $\sigma$, recorded at $t_\text{max}$, the end of the simulation. The system exhibits a transition from a disordered quasi-elastic regime at large $\gamma$ and $\alpha$ to an ordered collectively moving polar phase at small $\gamma$ and $\alpha$.
The existence of the transition is confirmed by a finite-size scaling analysis of the transition time $t^*$, defined as the time above which the polarization $\sigma>0.1$ [Fig.~\ref{fig:phasediag}(b)]. The later diverges as a power law for  a finite $\alpha_c(\gamma)<1$, the value of which decreases away from $1$ when the system size increases.
Our result contradicts a recent observation made for the dynamics of interacting dissipative active particles~\cite{Manacorda2017}. Deriving and solving kinetic equations, the authors show that, in the absence of noise, -- remember that tumbling is absent from our model -- the disordered state is always unstable. These results however strongly relies on the assumption that the velocity distributions are Gaussian. In the present case (see plots in the Supp. Mat.~\ref{suppmat}), the distributions have a highly non Gaussian shape; they exhibit a complex structure originating from the interplay between dissipation and activity. We believe such a qualitative difference can be responsible for this apparent contradiction.
A further argument in favor of the stability of the disordered phase for $\alpha<1$ runs as follow. Suppose a fluctuation allows for the formation of a train of $n$ particles. For simplicity, we assume the particles inside the train are regularly spaced by a distance $d<1/\phi$. In order to know whether this train grows, we compute the conditions under which it adsorbs a particle coming in the opposite direction (see Supp. Mat.~\ref{suppmat}). In the limit of $n\to\infty$, adsorption takes place when $\gamma<\gamma_c$ with:
\begin{equation}
\label{Gamma_infinity_main}
\gamma_c = -2 \frac{v_0}{d} \ln \frac{2\alpha}{1+\alpha}-\frac{v_0}{d}\frac{1-\alpha}{1+\alpha}.
\end{equation}
The corresponding curve with $d=1$, above which even infinitely long trains do not grow is plotted on top of the phase diagram [Fig. \ref{fig:phasediag}(c)]. For low $\alpha$, $\gamma_c$ underestimates the location of the transition because the velocity fluctuations within the train are likely to be very important. But for large enough $\alpha$, it provides a very good estimate of the transition location, and therefore clearly points at the existence of a finite size region of the parameter space where collective alignment does not take place. 
\begin{figure}[t!]
\vspace{-0mm}
\includegraphics[width=0.98\columnwidth,trim = 0mm 0mm 0mm 0mm, clip]{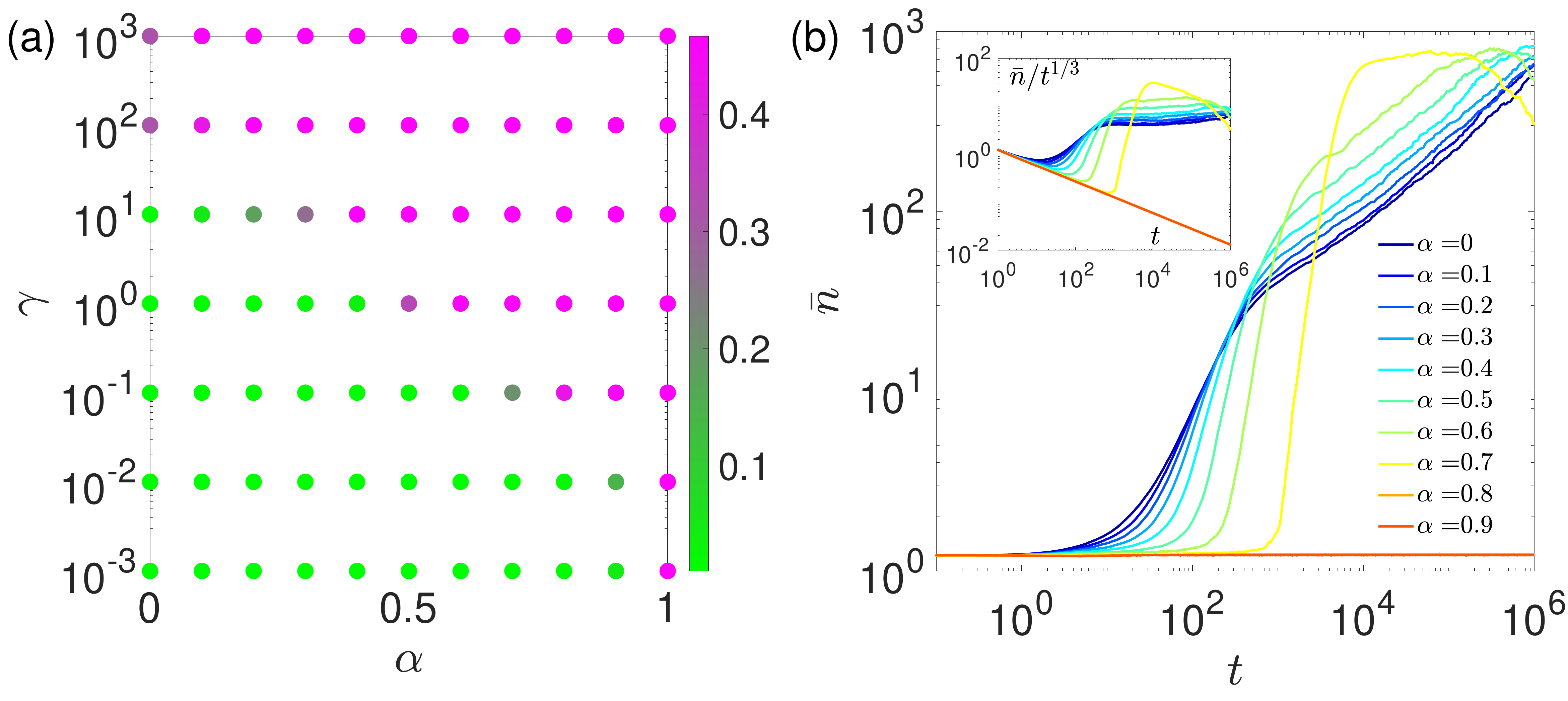}
\vspace{-0mm}
\caption{{\bf Transient clogs.} (a) Parameter space $(\alpha, \gamma)$ color coded by the minimal participation ratio $r_{\text{min}}$ observed during the simulation, averaged over 100 realizations, and $\phi=0.1$ (100 particles on 1000 sites). (b) Mean cluster size  $\bar{n}$ vs. time, for $\gamma=0.1$ ($10^3$ particles on a lattice of $10^4$ sites, average over 100 realizations). Inset: $\bar{n}$ rescaled by $t^{1/3}$. }
\label{fig:clog}
\vspace{-3mm}
\end{figure}

We finally discuss the transient condensation observed at low values of $\gamma$ and $\alpha$  [Fig. \ref{fig:dstnum}(d)]. [Fig. \ref{fig:clog}(a)] presents the phase diagram of the transient states, as obtained from the minimal value of the participation ratio $r_{\text{min}}$ recorded during the entire simulation. For low enough $\gamma$ and $\alpha$, the transition towards collective motion always takes place via transient clogs. As observed on numerical trajectories, a cluster, composed of two domains with opposite alignment facing each other, remains at rest for a typical time $\tau$ until the boundary between the two domains eventually diffuses up to an edge of the cluster [Fig. \ref{fig:clog}(b)]. The latter is then free to travel in the form of a train until it collides with the nearest arrested cluster or train, and so on. The growth rate of the polar ordering is then dominated by the slow diffusion process inside the clogs. The mass-weighted average cluster size~\cite{Stauffer1994}:
 \begin{align}
 \bar{n} =  \frac{\sum_n   n w_n}{\sum_n w_n}.
 \end{align}
with $w_n$ the mass fraction in clusters of size $n$ is indeed observed to grow as a weak power law, $\bar{n} \sim t^{1/3}$, before it saturates to the system size [Fig. \ref{fig:clog}(b)]. This can be understood within the context of a simple mean-field approximation. Let $K(m,n)$ be the rate at which two clusters of  size $m$ and $n$ aggregate to form a cluster of size $m+n$. Assuming that $K(am,an) \sim a^\lambda K(m,n)$, the typical cluster size obeys $\dot{\bar{n}} \sim \bar{n}^\lambda$, up to some proportionality constant \cite{Leyvraz2002, Krapivsky2010}. When $\lambda<1$, this yields $ \bar{n} \sim t^{1/(1-\lambda)}$. For a domain boundary diffusing as a random walker on a lattice, the mean first-passage time would scale as $\bar{n}^2$ \cite{Hughes1995,Redner2001}. Therefore, $\lambda=2$ and $\bar{n} \sim t^{1/3}$.

Altogether, \odbis{inspired by the intriguing observation of the onset of collective motion in a 1D system of swimming droplets, we have proposed a minimalistic model of active inelastic particles, in which the local alignment emerges from the combination of the spontaneous dispersion of speeds and absence of Galilean invariance.} We have shown that such local dynamical rules, are sufficient to induce a large scale transition to collective motion in 1D. The underlying mechanism could also be at play in higher dimension. Our results thus call for the development of a kinetic theory of active particles taking into account velocity fluctuations beyond the gaussian approximation~\cite{Manacorda2017}. Identifying a proper Ansatz for the velocity distribution is certainly one of the most challenging issue in the field of active liquids. The distributions reported here provide a first step in this direction.

{\it Acknowledgements.---} 
The authors acknowledge Vincent Démery for discussions about the alignement mechanism. P.I. acknowledges financial support from the People Programme (Marie Curie Actions) of the European Union’s Seventh Framework Programme (FP7/2007-2013) under REA grant agreement n. PCOFUND-GA-2013-609102, through the PRESTIGE programme coordinated by Campus France.

\vspace{-2mm}
\bibliographystyle{apsrev4-1}
\bibliography{Active.bib,1D.bib}

\begin{thebibliography}{39}%
\makeatletter
\providecommand \@ifxundefined [1]{%
 \@ifx{#1\undefined}
}%
\providecommand \@ifnum [1]{%
 \ifnum #1\expandafter \@firstoftwo
 \else \expandafter \@secondoftwo
 \fi
}%
\providecommand \@ifx [1]{%
 \ifx #1\expandafter \@firstoftwo
 \else \expandafter \@secondoftwo
 \fi
}%
\providecommand \natexlab [1]{#1}%
\providecommand \enquote  [1]{``#1''}%
\providecommand \bibnamefont  [1]{#1}%
\providecommand \bibfnamefont [1]{#1}%
\providecommand \citenamefont [1]{#1}%
\providecommand \href@noop [0]{\@secondoftwo}%
\providecommand \href [0]{\begingroup \@sanitize@url \@href}%
\providecommand \@href[1]{\@@startlink{#1}\@@href}%
\providecommand \@@href[1]{\endgroup#1\@@endlink}%
\providecommand \@sanitize@url [0]{\catcode `\\12\catcode `\$12\catcode
  `\&12\catcode `\#12\catcode `\^12\catcode `\_12\catcode `\%12\relax}%
\providecommand \@@startlink[1]{}%
\providecommand \@@endlink[0]{}%
\providecommand \url  [0]{\begingroup\@sanitize@url \@url }%
\providecommand \@url [1]{\endgroup\@href {#1}{\urlprefix }}%
\providecommand \urlprefix  [0]{URL }%
\providecommand \Eprint [0]{\href }%
\providecommand \doibase [0]{http://dx.doi.org/}%
\providecommand \selectlanguage [0]{\@gobble}%
\providecommand \bibinfo  [0]{\@secondoftwo}%
\providecommand \bibfield  [0]{\@secondoftwo}%
\providecommand \translation [1]{[#1]}%
\providecommand \BibitemOpen [0]{}%
\providecommand \bibitemStop [0]{}%
\providecommand \bibitemNoStop [0]{.\EOS\space}%
\providecommand \EOS [0]{\spacefactor3000\relax}%
\providecommand \BibitemShut  [1]{\csname bibitem#1\endcsname}%
\let\auto@bib@innerbib\@empty

\bibitem [{\citenamefont {Vicsek}\ and\ \citenamefont
  {Zafeiris}(2012)}]{Vicsek:2012ty}%
  \BibitemOpen
  \bibfield  {author} {\bibinfo {author} {\bibfnamefont {T.}~\bibnamefont
  {Vicsek}}\ and\ \bibinfo {author} {\bibfnamefont {A.}~\bibnamefont
  {Zafeiris}},\ }\href@noop {} {\bibfield  {journal} {\bibinfo  {journal}
  {Physics Reports}\ }\textbf {\bibinfo {volume} {517}},\ \bibinfo {pages} {71}
  (\bibinfo {year} {2012})}\BibitemShut {NoStop}%
\bibitem [{\citenamefont {Marchetti}\ \emph {et~al.}(2013)\citenamefont
  {Marchetti}, \citenamefont {Joanny}, \citenamefont {Ramaswamy}, \citenamefont
  {Liverpool}, \citenamefont {Prost}, \citenamefont {Rao},\ and\ \citenamefont
  {Simha}}]{Marchetti:2013bp}%
  \BibitemOpen
  \bibfield  {author} {\bibinfo {author} {\bibfnamefont {M.~C.}\ \bibnamefont
  {Marchetti}}, \bibinfo {author} {\bibfnamefont {J.-F.}\ \bibnamefont
  {Joanny}}, \bibinfo {author} {\bibfnamefont {S.}~\bibnamefont {Ramaswamy}},
  \bibinfo {author} {\bibfnamefont {T.~B.}\ \bibnamefont {Liverpool}}, \bibinfo
  {author} {\bibfnamefont {J.}~\bibnamefont {Prost}}, \bibinfo {author}
  {\bibfnamefont {M.}~\bibnamefont {Rao}}, \ and\ \bibinfo {author}
  {\bibfnamefont {R.~A.}\ \bibnamefont {Simha}},\ }\href@noop {} {\bibfield
  {journal} {\bibinfo  {journal} {Rev. Mod. Phys.}\ }\textbf {\bibinfo {volume}
  {85}},\ \bibinfo {pages} {1143} (\bibinfo {year} {2013})}\BibitemShut
  {NoStop}%
\bibitem [{\citenamefont {Bechinger}\ \emph {et~al.}(2016)\citenamefont
  {Bechinger}, \citenamefont {Di~Leonardo}, \citenamefont {L{\"o}wen},
  \citenamefont {Reichhardt}, \citenamefont {Volpe},\ and\ \citenamefont
  {Volpe}}]{Bechinger:2016cf}%
  \BibitemOpen
  \bibfield  {author} {\bibinfo {author} {\bibfnamefont {C.}~\bibnamefont
  {Bechinger}}, \bibinfo {author} {\bibfnamefont {R.}~\bibnamefont
  {Di~Leonardo}}, \bibinfo {author} {\bibfnamefont {H.}~\bibnamefont
  {L{\"o}wen}}, \bibinfo {author} {\bibfnamefont {C.}~\bibnamefont
  {Reichhardt}}, \bibinfo {author} {\bibfnamefont {G.}~\bibnamefont {Volpe}}, \
  and\ \bibinfo {author} {\bibfnamefont {G.}~\bibnamefont {Volpe}},\
  }\href@noop {} {\bibfield  {journal} {\bibinfo  {journal} {Rev. Mod. Phys.}\
  }\textbf {\bibinfo {volume} {88}},\ \bibinfo {pages} {1} (\bibinfo {year}
  {2016})}\BibitemShut {NoStop}%
\bibitem [{\citenamefont {Chat{\'e}}\ \emph {et~al.}(2008)\citenamefont
  {Chat{\'e}}, \citenamefont {Ginelli}, \citenamefont {Gr{\'e}goire},\ and\
  \citenamefont {Raynaud}}]{Chate:2008isb}%
  \BibitemOpen
  \bibfield  {author} {\bibinfo {author} {\bibfnamefont {H.}~\bibnamefont
  {Chat{\'e}}}, \bibinfo {author} {\bibfnamefont {F.}~\bibnamefont {Ginelli}},
  \bibinfo {author} {\bibfnamefont {G.}~\bibnamefont {Gr{\'e}goire}}, \ and\
  \bibinfo {author} {\bibfnamefont {F.}~\bibnamefont {Raynaud}},\ }\href@noop
  {} {\bibfield  {journal} {\bibinfo  {journal} {Phys. Rev. E}\ }\textbf
  {\bibinfo {volume} {77}},\ \bibinfo {pages} {046113} (\bibinfo {year}
  {2008})}\BibitemShut {NoStop}%
\bibitem [{\citenamefont {Solon}\ and\ \citenamefont
  {Tailleur}(2013)}]{Solon:2013vr}%
  \BibitemOpen
  \bibfield  {author} {\bibinfo {author} {\bibfnamefont {A.~P.}\ \bibnamefont
  {Solon}}\ and\ \bibinfo {author} {\bibfnamefont {J.}~\bibnamefont
  {Tailleur}},\ }\href@noop {} {\bibfield  {journal} {\bibinfo  {journal}
  {PRL}\ }\textbf {\bibinfo {volume} {111}},\ \bibinfo {pages} {1} (\bibinfo
  {year} {2013})}\BibitemShut {NoStop}%
\bibitem [{\citenamefont {Peshkov}\ \emph {et~al.}(2014)\citenamefont
  {Peshkov}, \citenamefont {Bertin}, \citenamefont {Ginelli},\ and\
  \citenamefont {Chat{\'e}}}]{Peshkov:2014un}%
  \BibitemOpen
  \bibfield  {author} {\bibinfo {author} {\bibfnamefont {A.}~\bibnamefont
  {Peshkov}}, \bibinfo {author} {\bibfnamefont {E.~M.}\ \bibnamefont {Bertin}},
  \bibinfo {author} {\bibfnamefont {F.}~\bibnamefont {Ginelli}}, \ and\
  \bibinfo {author} {\bibfnamefont {H.}~\bibnamefont {Chat{\'e}}},\ }\href@noop
  {} {\bibfield  {journal} {\bibinfo  {journal} {The European Physical
  Journal-Special Topics}\ }\textbf {\bibinfo {volume} {223}},\ \bibinfo
  {pages} {1315} (\bibinfo {year} {2014})}\BibitemShut {NoStop}%
\bibitem [{\citenamefont {Cates}\ and\ \citenamefont
  {Tailleur}(2015)}]{Cates:2015ft}%
  \BibitemOpen
  \bibfield  {author} {\bibinfo {author} {\bibfnamefont {M.~E.}\ \bibnamefont
  {Cates}}\ and\ \bibinfo {author} {\bibfnamefont {J.}~\bibnamefont
  {Tailleur}},\ }\href@noop {} {\bibfield  {journal} {\bibinfo  {journal}
  {Annual Review of Condensed Matter Physics}\ }\textbf {\bibinfo {volume}
  {6}},\ \bibinfo {pages} {219} (\bibinfo {year} {2015})}\BibitemShut {NoStop}%
\bibitem [{\citenamefont {Farrell}\ \emph {et~al.}(2012)\citenamefont
  {Farrell}, \citenamefont {Marchetti}, \citenamefont {Marenduzzo},\ and\
  \citenamefont {Tailleur}}]{Farrell:2012ks}%
  \BibitemOpen
  \bibfield  {author} {\bibinfo {author} {\bibfnamefont {F.~D.~C.}\
  \bibnamefont {Farrell}}, \bibinfo {author} {\bibfnamefont {M.~C.}\
  \bibnamefont {Marchetti}}, \bibinfo {author} {\bibfnamefont {D.}~\bibnamefont
  {Marenduzzo}}, \ and\ \bibinfo {author} {\bibfnamefont {J.}~\bibnamefont
  {Tailleur}},\ }\href@noop {} {\bibfield  {journal} {\bibinfo  {journal}
  {PRL}\ }\textbf {\bibinfo {volume} {108}},\ \bibinfo {pages} {248101}
  (\bibinfo {year} {2012})}\BibitemShut {NoStop}%
\bibitem [{\citenamefont {Sese-Sansa}\ \emph {et~al.}(2018)\citenamefont
  {Sese-Sansa}, \citenamefont {Pagonabarraga},\ and\ \citenamefont
  {Levis}}]{SeseSansa:2018wl}%
  \BibitemOpen
  \bibfield  {author} {\bibinfo {author} {\bibfnamefont {E.}~\bibnamefont
  {Sese-Sansa}}, \bibinfo {author} {\bibfnamefont {I.}~\bibnamefont
  {Pagonabarraga}}, \ and\ \bibinfo {author} {\bibfnamefont {D.}~\bibnamefont
  {Levis}},\ }\href@noop {} {\bibfield  {journal} {\bibinfo  {journal} {arXiv}\
  } (\bibinfo {year} {2018})},\ \Eprint {http://arxiv.org/abs/1807.07497v1}
  {1807.07497v1} \BibitemShut {NoStop}%
\bibitem [{\citenamefont {Shi}\ and\ \citenamefont
  {Chat{\'e}}(2018)}]{Shi:2018tp}%
  \BibitemOpen
  \bibfield  {author} {\bibinfo {author} {\bibfnamefont {X.-q.}\ \bibnamefont
  {Shi}}\ and\ \bibinfo {author} {\bibfnamefont {H.}~\bibnamefont
  {Chat{\'e}}},\ }\href@noop {} {\bibfield  {journal} {\bibinfo  {journal}
  {arXiv}\ } (\bibinfo {year} {2018})},\ \Eprint
  {http://arxiv.org/abs/1807.00294v2} {1807.00294v2} \BibitemShut {NoStop}%
\bibitem [{\citenamefont {van~der Linden}\ \emph {et~al.}(2019)\citenamefont
  {van~der Linden}, \citenamefont {Alexander}, \citenamefont {Aarts},\ and\
  \citenamefont {Dauchot}}]{vanderLinden:2019kd}%
  \BibitemOpen
  \bibfield  {author} {\bibinfo {author} {\bibfnamefont {M.~N.}\ \bibnamefont
  {van~der Linden}}, \bibinfo {author} {\bibfnamefont {L.~C.}\ \bibnamefont
  {Alexander}}, \bibinfo {author} {\bibfnamefont {D.~G. A.~L.}\ \bibnamefont
  {Aarts}}, \ and\ \bibinfo {author} {\bibfnamefont {O.}~\bibnamefont
  {Dauchot}},\ }\href@noop {} {\bibfield  {journal} {\bibinfo  {journal} {Phys.
  Rev. Lett.}\ }\textbf {\bibinfo {volume} {123}},\ \bibinfo {pages} {098001}
  (\bibinfo {year} {2019})}\BibitemShut {NoStop}%
\bibitem [{\citenamefont {Lozano}\ \emph {et~al.}(2019)\citenamefont {Lozano},
  \citenamefont {Gomez-Solano},\ and\ \citenamefont
  {Bechinger}}]{Lozano:2019er}%
  \BibitemOpen
  \bibfield  {author} {\bibinfo {author} {\bibfnamefont {C.}~\bibnamefont
  {Lozano}}, \bibinfo {author} {\bibfnamefont {J.~R.}\ \bibnamefont
  {Gomez-Solano}}, \ and\ \bibinfo {author} {\bibfnamefont {C.}~\bibnamefont
  {Bechinger}},\ }\href@noop {} {\bibfield  {journal} {\bibinfo  {journal}
  {Nature Materials}\ ,\ \bibinfo {pages} {1}} (\bibinfo {year}
  {2019})}\BibitemShut {NoStop}%
\bibitem [{\citenamefont {Biondi}\ \emph {et~al.}()\citenamefont {Biondi},
  \citenamefont {Quinn}, \citenamefont {Journal},\ and\ \citenamefont
  {{1998}}}]{Biondi:up}%
  \BibitemOpen
  \bibfield  {author} {\bibinfo {author} {\bibfnamefont {S.~A.}\ \bibnamefont
  {Biondi}}, \bibinfo {author} {\bibfnamefont {J.~A.}\ \bibnamefont {Quinn}},
  \bibinfo {author} {\bibfnamefont {H.~G.~A.}\ \bibnamefont {Journal}}, \ and\
  \bibinfo {author} {\bibnamefont {{1998}}},\ }\href@noop {} {\bibinfo
  {journal} {Wiley Online Library}\ }\BibitemShut {NoStop}%
\bibitem [{\citenamefont {Jeli{\'{c}}}\ \emph {et~al.}(2012)\citenamefont
  {Jeli{\'{c}}}, \citenamefont {Appert-Rolland}, \citenamefont {Lemercier},\
  and\ \citenamefont {Pettr{\'{e}}}}]{Jelic2012}%
  \BibitemOpen
\bibfield  {journal} {  }\bibfield  {author} {\bibinfo {author} {\bibfnamefont
  {A.}~\bibnamefont {Jeli{\'{c}}}}, \bibinfo {author} {\bibfnamefont
  {C.}~\bibnamefont {Appert-Rolland}}, \bibinfo {author} {\bibfnamefont
  {S.}~\bibnamefont {Lemercier}}, \ and\ \bibinfo {author} {\bibfnamefont
  {J.}~\bibnamefont {Pettr{\'{e}}}},\ }\href@noop {} {\bibfield  {journal}
  {\bibinfo  {journal} {Phys. Rev. E}\ }\textbf {\bibinfo {volume} {85}},\
  \bibinfo {pages} {036111} (\bibinfo {year} {2012})}\BibitemShut {NoStop}%
\bibitem [{\citenamefont {Chou}\ \emph {et~al.}(2011)\citenamefont {Chou},
  \citenamefont {Mallick},\ and\ \citenamefont {Zia}}]{Chou2011}%
  \BibitemOpen
  \bibfield  {author} {\bibinfo {author} {\bibfnamefont {T.}~\bibnamefont
  {Chou}}, \bibinfo {author} {\bibfnamefont {K.}~\bibnamefont {Mallick}}, \
  and\ \bibinfo {author} {\bibfnamefont {R.~K.~P.}\ \bibnamefont {Zia}},\
  }\href@noop {} {\bibfield  {journal} {\bibinfo  {journal} {Reports on
  Progress in Physics}\ }\textbf {\bibinfo {volume} {74}},\ \bibinfo {pages}
  {116601} (\bibinfo {year} {2011})}\BibitemShut {NoStop}%
\bibitem [{\citenamefont {Lieb}\ and\ \citenamefont
  {Mattis}(2013)}]{lieb2013mathematical}%
  \BibitemOpen
  \bibfield  {author} {\bibinfo {author} {\bibfnamefont {E.~H.}\ \bibnamefont
  {Lieb}}\ and\ \bibinfo {author} {\bibfnamefont {D.~C.}\ \bibnamefont
  {Mattis}},\ }\href@noop {} {\emph {\bibinfo {title} {Mathematical physics in
  one dimension: exactly soluble models of interacting particles}}}\ (\bibinfo
  {publisher} {Academic Press},\ \bibinfo {year} {2013})\BibitemShut {NoStop}%
\bibitem [{\citenamefont {Privman}(2005)}]{privman2005nonequilibrium}%
  \BibitemOpen
  \bibfield  {author} {\bibinfo {author} {\bibfnamefont {V.}~\bibnamefont
  {Privman}},\ }\href@noop {} {\emph {\bibinfo {title} {Nonequilibrium
  statistical mechanics in one dimension}}}\ (\bibinfo  {publisher} {Cambridge
  University Press},\ \bibinfo {year} {2005})\BibitemShut {NoStop}%
\bibitem [{\citenamefont {Arratia}(1983)}]{Arratia1983}%
  \BibitemOpen
  \bibfield  {author} {\bibinfo {author} {\bibfnamefont {R.}~\bibnamefont
  {Arratia}},\ }\href@noop {} {\bibfield  {journal} {\bibinfo  {journal} {The
  Annals of Probability}\ }\textbf {\bibinfo {volume} {11}},\ \bibinfo {pages}
  {362} (\bibinfo {year} {1983})}\BibitemShut {NoStop}%
\bibitem [{\citenamefont {Harris}(1965)}]{Harris1965}%
  \BibitemOpen
  \bibfield  {author} {\bibinfo {author} {\bibfnamefont {T.}~\bibnamefont
  {Harris}},\ }\href@noop {} {\bibfield  {journal} {\bibinfo  {journal}
  {Journal of Applied Probability}\ }\textbf {\bibinfo {volume} {2}},\ \bibinfo
  {pages} {323} (\bibinfo {year} {1965})}\BibitemShut {NoStop}%
\bibitem [{\citenamefont {Du}\ \emph {et~al.}(1995)\citenamefont {Du},
  \citenamefont {Li},\ and\ \citenamefont {{Kadanoff, Leo P.}}}]{Du:1995kw}%
  \BibitemOpen
  \bibfield  {author} {\bibinfo {author} {\bibfnamefont {Y.}~\bibnamefont
  {Du}}, \bibinfo {author} {\bibfnamefont {H.}~\bibnamefont {Li}}, \ and\
  \bibinfo {author} {\bibnamefont {{Kadanoff, Leo P.}}},\ }\href@noop {}
  {\bibfield  {journal} {\bibinfo  {journal} {Phys. Rev. Lett.}\ }\textbf
  {\bibinfo {volume} {74}},\ \bibinfo {pages} {1268} (\bibinfo {year}
  {1995})}\BibitemShut {NoStop}%
\bibitem [{\citenamefont {Ben-Naim}\ \emph {et~al.}(1999)\citenamefont
  {Ben-Naim}, \citenamefont {Chen}, \citenamefont {Doolen},\ and\ \citenamefont
  {Redner}}]{BenNaim:1999gj}%
  \BibitemOpen
  \bibfield  {author} {\bibinfo {author} {\bibfnamefont {E.}~\bibnamefont
  {Ben-Naim}}, \bibinfo {author} {\bibfnamefont {S.~Y.}\ \bibnamefont {Chen}},
  \bibinfo {author} {\bibfnamefont {G.~D.}\ \bibnamefont {Doolen}}, \ and\
  \bibinfo {author} {\bibfnamefont {S.}~\bibnamefont {Redner}},\ }\href@noop {}
  {\bibfield  {journal} {\bibinfo  {journal} {Phys. Rev. Lett.}\ }\textbf
  {\bibinfo {volume} {83}},\ \bibinfo {pages} {4069} (\bibinfo {year}
  {1999})}\BibitemShut {NoStop}%
\bibitem [{\citenamefont {Baldassarri}\ \emph {et~al.}(2018)\citenamefont
  {Baldassarri}, \citenamefont {Puglisi},\ and\ \citenamefont
  {Prados}}]{Baldassarri:2018id}%
  \BibitemOpen
  \bibfield  {author} {\bibinfo {author} {\bibfnamefont {A.}~\bibnamefont
  {Baldassarri}}, \bibinfo {author} {\bibfnamefont {A.}~\bibnamefont
  {Puglisi}}, \ and\ \bibinfo {author} {\bibfnamefont {A.}~\bibnamefont
  {Prados}},\ }\href@noop {} {\ ,\ \bibinfo {pages} {1} (\bibinfo {year}
  {2018})}\BibitemShut {NoStop}%
\bibitem [{\citenamefont {Mallick}(2015)}]{Mallick2015}%
  \BibitemOpen
  \bibfield  {author} {\bibinfo {author} {\bibfnamefont {K.}~\bibnamefont
  {Mallick}},\ }\href@noop {} {\bibfield  {journal} {\bibinfo  {journal}
  {Physica A}\ }\textbf {\bibinfo {volume} {418}},\ \bibinfo {pages} {17}
  (\bibinfo {year} {2015})}\BibitemShut {NoStop}%
\bibitem [{\citenamefont {Manacorda}\ and\ \citenamefont
  {Puglisi}(2017{\natexlab{a}})}]{Manacorda:2017eia}%
  \BibitemOpen
  \bibfield  {author} {\bibinfo {author} {\bibfnamefont {A.}~\bibnamefont
  {Manacorda}}\ and\ \bibinfo {author} {\bibfnamefont {A.}~\bibnamefont
  {Puglisi}},\ }\href@noop {} {\bibfield  {journal} {\bibinfo  {journal} {PRL}\
  }\textbf {\bibinfo {volume} {119}},\ \bibinfo {pages} {208003} (\bibinfo
  {year} {2017}{\natexlab{a}})}\BibitemShut {NoStop}%
\bibitem [{\citenamefont {Nguyen Thu~Lam}\ \emph {et~al.}(2015)\citenamefont
  {Nguyen Thu~Lam}, \citenamefont {Schindler},\ and\ \citenamefont
  {Dauchot}}]{Lam:2015jr}%
  \BibitemOpen
  \bibfield  {author} {\bibinfo {author} {\bibfnamefont {K.-D.}\ \bibnamefont
  {Nguyen Thu~Lam}}, \bibinfo {author} {\bibfnamefont {M.}~\bibnamefont
  {Schindler}}, \ and\ \bibinfo {author} {\bibfnamefont {O.}~\bibnamefont
  {Dauchot}},\ }\href@noop {} {\bibfield  {journal} {\bibinfo  {journal} {J.
  Stat. Mech.}\ }\textbf {\bibinfo {volume} {2015}},\ \bibinfo {pages} {P10017}
  (\bibinfo {year} {2015})}\BibitemShut {NoStop}%
\bibitem [{\citenamefont {Solon}\ and\ \citenamefont
  {Tailleur}(2015)}]{Solon:2015he}%
  \BibitemOpen
  \bibfield  {author} {\bibinfo {author} {\bibfnamefont {A.~P.}\ \bibnamefont
  {Solon}}\ and\ \bibinfo {author} {\bibfnamefont {J.}~\bibnamefont
  {Tailleur}},\ }\href@noop {} {\bibfield  {journal} {\bibinfo  {journal}
  {Phys. Rev. E}\ }\textbf {\bibinfo {volume} {92}},\ \bibinfo {pages} {042119}
  (\bibinfo {year} {2015})}\BibitemShut {NoStop}%
\bibitem [{\citenamefont {Soto}\ and\ \citenamefont
  {Golestanian}(2014)}]{Soto2014}%
  \BibitemOpen
  \bibfield  {author} {\bibinfo {author} {\bibfnamefont {R.}~\bibnamefont
  {Soto}}\ and\ \bibinfo {author} {\bibfnamefont {R.}~\bibnamefont
  {Golestanian}},\ }\href@noop {} {\bibfield  {journal} {\bibinfo  {journal}
  {Physical Review Letters}\ }\textbf {\bibinfo {volume} {112}},\ \bibinfo
  {pages} {068301} (\bibinfo {year} {2014})}\BibitemShut {NoStop}%
\bibitem [{\citenamefont {Slowman}\ \emph {et~al.}(2016)\citenamefont
  {Slowman}, \citenamefont {Evans},\ and\ \citenamefont
  {Blythe}}]{Slowman2016}%
  \BibitemOpen
  \bibfield  {author} {\bibinfo {author} {\bibfnamefont {A.~B.}\ \bibnamefont
  {Slowman}}, \bibinfo {author} {\bibfnamefont {M.~R.}\ \bibnamefont {Evans}},
  \ and\ \bibinfo {author} {\bibfnamefont {R.~A.}\ \bibnamefont {Blythe}},\
  }\href@noop {} {\bibfield  {journal} {\bibinfo  {journal} {Phys. Rev. Lett.}\
  }\textbf {\bibinfo {volume} {116}},\ \bibinfo {pages} {218101} (\bibinfo
  {year} {2016})}\BibitemShut {NoStop}%
\bibitem [{\citenamefont {Slowman}\ \emph {et~al.}(2017)\citenamefont
  {Slowman}, \citenamefont {Evans},\ and\ \citenamefont
  {Blythe}}]{Slowman2017}%
  \BibitemOpen
  \bibfield  {author} {\bibinfo {author} {\bibfnamefont {A.~B.}\ \bibnamefont
  {Slowman}}, \bibinfo {author} {\bibfnamefont {M.~R.}\ \bibnamefont {Evans}},
  \ and\ \bibinfo {author} {\bibfnamefont {R.~A.}\ \bibnamefont {Blythe}},\
  }\href@noop {} {\bibfield  {journal} {\bibinfo  {journal} {J. Phys. A}\
  }\textbf {\bibinfo {volume} {50}},\ \bibinfo {pages} {375601} (\bibinfo
  {year} {2017})}\BibitemShut {NoStop}%
\bibitem [{\citenamefont {Locatelli}\ \emph {et~al.}(2015)\citenamefont
  {Locatelli}, \citenamefont {Baldovin}, \citenamefont {Orlandini},\ and\
  \citenamefont {Pierno}}]{Locatelli:2015it}%
  \BibitemOpen
  \bibfield  {author} {\bibinfo {author} {\bibfnamefont {E.}~\bibnamefont
  {Locatelli}}, \bibinfo {author} {\bibfnamefont {F.}~\bibnamefont {Baldovin}},
  \bibinfo {author} {\bibfnamefont {E.}~\bibnamefont {Orlandini}}, \ and\
  \bibinfo {author} {\bibfnamefont {M.}~\bibnamefont {Pierno}},\ }\href@noop {}
  {\bibfield  {journal} {\bibinfo  {journal} {Phys. Rev. E}\ }\textbf {\bibinfo
  {volume} {91}},\ \bibinfo {pages} {1428} (\bibinfo {year}
  {2015})}\BibitemShut {NoStop}%
\bibitem [{\citenamefont {Kourbane-Houssene}\ \emph {et~al.}(2018)\citenamefont
  {Kourbane-Houssene}, \citenamefont {Erignoux}, \citenamefont {Bodineau},\
  and\ \citenamefont {Tailleur}}]{Kourbane-Houssene2018a}%
  \BibitemOpen
  \bibfield  {author} {\bibinfo {author} {\bibfnamefont {M.}~\bibnamefont
  {Kourbane-Houssene}}, \bibinfo {author} {\bibfnamefont {C.}~\bibnamefont
  {Erignoux}}, \bibinfo {author} {\bibfnamefont {T.}~\bibnamefont {Bodineau}},
  \ and\ \bibinfo {author} {\bibfnamefont {J.}~\bibnamefont {Tailleur}},\
  }\href@noop {} {\bibfield  {journal} {\bibinfo  {journal} {Phys. Rev. Lett.}\
  }\textbf {\bibinfo {volume} {120}},\ \bibinfo {pages} {268003} (\bibinfo
  {year} {2018})}\BibitemShut {NoStop}%
\bibitem [{\citenamefont {Izri}\ \emph {et~al.}(2014)\citenamefont {Izri},
  \citenamefont {van~der Linden}, \citenamefont {Michelin},\ and\ \citenamefont
  {Dauchot}}]{Izri:2014fv}%
  \BibitemOpen
  \bibfield  {author} {\bibinfo {author} {\bibfnamefont {Z.}~\bibnamefont
  {Izri}}, \bibinfo {author} {\bibfnamefont {M.~N.}\ \bibnamefont {van~der
  Linden}}, \bibinfo {author} {\bibfnamefont {S.}~\bibnamefont {Michelin}}, \
  and\ \bibinfo {author} {\bibfnamefont {O.}~\bibnamefont {Dauchot}},\
  }\href@noop {} {\bibfield  {journal} {\bibinfo  {journal} {Phys. Rev. Lett.}\
  }\textbf {\bibinfo {volume} {113}},\ \bibinfo {pages} {248302} (\bibinfo
  {year} {2014})}\BibitemShut {NoStop}%
\bibitem [{\citenamefont {Michelin}\ \emph {et~al.}(2013)\citenamefont
  {Michelin}, \citenamefont {Lauga},\ and\ \citenamefont
  {Bartolo}}]{Michelin:2013gv}%
  \BibitemOpen
  \bibfield  {author} {\bibinfo {author} {\bibfnamefont {S.}~\bibnamefont
  {Michelin}}, \bibinfo {author} {\bibfnamefont {E.}~\bibnamefont {Lauga}}, \
  and\ \bibinfo {author} {\bibfnamefont {D.}~\bibnamefont {Bartolo}},\
  }\href@noop {} {\bibfield  {journal} {\bibinfo  {journal} {Phys. Fluids}\
  }\textbf {\bibinfo {volume} {25}},\ \bibinfo {pages} {061701} (\bibinfo
  {year} {2013})}\BibitemShut {NoStop}%
\bibitem [{\citenamefont {Manacorda}\ and\ \citenamefont
  {Puglisi}(2017{\natexlab{b}})}]{Manacorda2017}%
  \BibitemOpen
  \bibfield  {author} {\bibinfo {author} {\bibfnamefont {A.}~\bibnamefont
  {Manacorda}}\ and\ \bibinfo {author} {\bibfnamefont {A.}~\bibnamefont
  {Puglisi}},\ }\href@noop {} {\bibfield  {journal} {\bibinfo  {journal}
  {Physical Review Letters}\ }\textbf {\bibinfo {volume} {119}},\ \bibinfo
  {pages} {1} (\bibinfo {year} {2017}{\natexlab{b}})}\BibitemShut {NoStop}%
\bibitem [{\citenamefont {Stauffer}\ and\ \citenamefont
  {Aharony}(1994)}]{Stauffer1994}%
  \BibitemOpen
  \bibfield  {author} {\bibinfo {author} {\bibfnamefont {D.}~\bibnamefont
  {Stauffer}}\ and\ \bibinfo {author} {\bibfnamefont {A.}~\bibnamefont
  {Aharony}},\ }\href@noop {} {\emph {\bibinfo {title} {{Introduction to
  Percolation Theory}}}}\ (\bibinfo  {publisher} {Taylor and Francis},\
  \bibinfo {year} {1994})\BibitemShut {NoStop}%
\bibitem [{\citenamefont {Leyvraz}\ and\ \citenamefont
  {Redner}(2002)}]{Leyvraz2002}%
  \BibitemOpen
  \bibfield  {author} {\bibinfo {author} {\bibfnamefont {F.}~\bibnamefont
  {Leyvraz}}\ and\ \bibinfo {author} {\bibfnamefont {S.}~\bibnamefont
  {Redner}},\ }\href@noop {} {\bibfield  {journal} {\bibinfo  {journal}
  {Physical Review Letters}\ }\textbf {\bibinfo {volume} {88}},\ \bibinfo
  {pages} {068301} (\bibinfo {year} {2002})}\BibitemShut {NoStop}%
\bibitem [{\citenamefont {Krapivsky}\ \emph {et~al.}(2010)\citenamefont
  {Krapivsky}, \citenamefont {Redner},\ and\ \citenamefont
  {Ben-Naim}}]{Krapivsky2010}%
  \BibitemOpen
  \bibfield  {author} {\bibinfo {author} {\bibfnamefont {P.~L.}\ \bibnamefont
  {Krapivsky}}, \bibinfo {author} {\bibfnamefont {S.}~\bibnamefont {Redner}}, \
  and\ \bibinfo {author} {\bibfnamefont {E.}~\bibnamefont {Ben-Naim}},\
  }\href@noop {} {\emph {\bibinfo {title} {{A Kinetic View of Statistical
  Physics}}}}\ (\bibinfo  {publisher} {Cambridge University Press},\ \bibinfo
  {year} {2010})\BibitemShut {NoStop}%
\bibitem [{\citenamefont {Hughes}(1995)}]{Hughes1995}%
  \BibitemOpen
  \bibfield  {author} {\bibinfo {author} {\bibfnamefont {B.~D.}\ \bibnamefont
  {Hughes}},\ }\href@noop {} {\emph {\bibinfo {title} {{Random Walks and Random
  Environments: Random walks, Volume 1}}}}\ (\bibinfo  {publisher} {Oxford
  University Press},\ \bibinfo {address} {Oxford},\ \bibinfo {year}
  {1995})\BibitemShut {NoStop}%
\bibitem [{\citenamefont {Redner}(2001)}]{Redner2001}%
  \BibitemOpen
  \bibfield  {author} {\bibinfo {author} {\bibfnamefont {S.}~\bibnamefont
  {Redner}},\ }\href@noop {} {\emph {\bibinfo {title} {{A Guide to
  First-Passage Processes}}}}\ (\bibinfo  {publisher} {Cambridge University
  Press},\ \bibinfo {address} {Cambridge},\ \bibinfo {year} {2001})\BibitemShut
  {NoStop}%
\bibitem []{suppmat}%
  \BibitemOpen
  \bibfield  {author} {\bibinfo {author} {\bibfnamefont {See Supplemental Material at [URL will be inserted by publisher] for a movie of the experiment, experimental methods, numerical velocity distributions and derivation of eq.[3]}},\ }\href@noop {} \BibitemShut {NoStop}%

\end{thebibliography}%

\end{document}